\documentclass[prl,preprintnumbers,twocolumn,floats,aps,epsfig,nofootinbib,axodraw,amsmath,amssymb]{revtex4}
\usepackage{subfigure}
\def\beq{\begin{equation}}
\def\eeq{\end{equation}}
\def\bea{\begin{eqnarray}}
\def\eea{\end{eqnarray}}

\def\etmiss{E\!\!\!\!\slash_{T}}
\def\pslash{\not{\hbox{\kern-4pt $p$}}}
\def\qslash{\not{\hbox{\kern-4pt $q$}}}
\def\lv{\not{\hbox{\kern-4pt $L$}}}
\def\lsim{\mathrel{\raise.3ex\hbox{$<$\kern-.75em\lower1ex\hbox{$\sim$}}}}
\def\gsim{\mathrel{\raise.3ex\hbox{$>$\kern-.75em\lower1ex\hbox{$\sim$}}}}
\def\ifmath#1{\relax\ifmmode #1\else $#1$\fi}

\usepackage{graphicx}
\usepackage{bm}
\begin{document}
\draft
\renewcommand{\thefootnote}{\arabic{footnote}}

\preprint{UCB-PTH-09/24}

\title{Dark Matter Stabilization Symmetries from \\ Spontaneous Symmetry Breaking}
\bigskip
\author{Devin G. E. Walker\footnote{Email Address:  dgwalker@berkeley.edu.}}
\address{Department of Physics, University of California, Berkeley, CA 94720, U.S.A,\\
Theoretical Physics Group, Lawrence Berkeley National Laboratory, Berkeley, CA 94720, U.S.A.}

\begin{abstract}
We present a class of models in which the dark matter stabilization symmetry is generated by spontaneous symmetry breaking.  These models naturally correlate the dark and electroweak symmetry breaking scales.  The result is a generic mechanism linking the annihilation cross section for thermally populated dark matter with the weak scale.  The thermal relic abundance, sensitivity to major precision electroweak observables and additional LHC signatures are also presented. 
\end{abstract}

\maketitle
The CERN Large Hadron Collider (LHC) is expected to begin taking the first steps toward probing the TeV energy scale by the fall of 2009.  Besides the long anticipated discovery of the Higgs boson and clues about the mechanism that stabilizes the weak scale, the LHC is expected to shed some light on the nature of dark matter.  All viable dark matter (DM) candidates must be stable, neutral under the Standard Model (SM), non-relativistic at redshifts of $z \sim 3000$ and generate the measured relic abundance of  $h^2\, \Omega_{\mathrm{DM}} = 0.1131 \pm 0.0034$~\cite{Hinshaw:2008kr}.  It is intriguing to note the correct thermal relic abundance is generated for annihilation cross sections that are typical for the weak interactions~\cite{Kolb:1990vq}
\begin{equation}
\Omega_\mathrm{DM} \,h^{2} \simeq \mathrm{constant}\cdot {T^3 \over M_\mathrm{pl}^3 \,\langle \sigma v \rangle}\simeq {0.1 \,\,\mathrm{pb} \cdot c\over \langle \sigma v \rangle}
\end{equation}
where $c$, $M_\mathrm{pl}$ and $T$ are the speed of light, Planck mass, and freeze out temperature, respectively.  Here the annihilation cross section, $\sigma$, for thermally populated dark matter, has an overall mass scale of order the weak scale.  This correlation is known as the ``WIMP miracle."\footnote{It has been noted that dark matter candidates can get the correct thermal relic abundance with the right ratio of couplings and overall mass scale; this is true even if the overall mass scale (coupling) is not the weak scale (weak strength).\cite{Feng:2008ya}} In this letter, we present a class of models in which the dark matter stabilization symmetry is a remnant of a spontaneously broken gauge symmetry.  One result is a natural mechanism which correlates the dark and electroweak symmetry breaking scales.  In two companion papers~\cite{Walker:2009ph,agashe:2009ph}, these models are shown to give distinct signatures at the LHC.  

Addressing the hierarchy problem has produced many models with viable DM candidates~\cite{Bertone:2004pz}.  Yet, the hierarchy problem is not necessarily related to the question of the origin of dark matter.  To motivate searching for an alternative, note that many of these popular models stabilize dark matter with a $\mathcal{Z}_2$ symmetry (generic parity).  Since particles are classified by their symmetries, only one type of candidate is being considered!  Consider the alternatives:  Dark matter can be stabilized by either discrete symmetries (such as  $\mathcal{Z}_2 \times \mathcal{Z}_2$, $\mathcal{Z}_3$~$\ldots$) and/or continuous global symmetries.  Continuous gauge symmetries alone cannot stabilize dark matter and must either confine or undergo spontaneous symmetry breaking.  To see this, consider a scenario with a gauged abelian stabilization symmetry.  Such a model would be problematic due to constraints from fifth force measurements, nucleosynthesis bounds, as well as potentially generate milli-electric charges for the dark matter via kinetic mixing of the new gauge boson with the photon~\cite{abelian}.  Gauged non-abelian stabilization symmetries  fare no better.  For temperatures below the mass of the dark matter candidate, such a model becomes an effective Yang-Mills theory and eventually strongly coupled.  This scenario leads to the DM candidates confining into SM singlets and decaying.\footnote{The exception is a gauge theory in a free infrared phase.  Those models would be subject to nucleosynthesis bounds because of the required near-massless fermions.}  Gauge theories can, however, provide motivation for the origin of dark matter stabilization symmetries.  DM stabilized by continuous global symmetries can be thought of as a low-energy effective description of a confining gauge theory where the QCD analogy of baryon number is preserved.\footnote{In coming work, the collider signatures for the strongly coupled scenarios are distinguished from perturbative scenarios in \cite{georgi:2009ph2}.}  Discrete symmetries can be the result of spontaneous breaking of perturbative gauge symmetries.  

In this letter, we explore the possibility that the dark matter is stabilized by an unbroken discrete subgroup of a spontaneously broken dark gauge symmetry.  Because of the strong connection between the weak scale and the annihilation cross section for thermally produced dark matter, we take the new dark gauge group to be broken at the electroweak scale.  This implies a new weak scale dark gauge boson, which could provide experimentally distinguishable signatures.~\cite{Walker:2009ph,agashe:2009ph}  The simplest scenario, spontaneously breaking a U(1), is potentially problematic.  A new U(1) gauge boson can lead to large precision electroweak corrections through kinetic mixing with SM hypercharge at tree level
\begin{equation}
\mathcal{O}_1 = B_{\mu\nu} F^{\mu\nu}.
\end{equation}
Here $B_{\mu\nu}$ and $F_{\mu\nu}$ are the field strength tensors for hypercharge and new U(1) gauge bosons, respectively.  This operator is valid for models where a charge conjugation symmetry in the hidden sector is broken; thus, a viable model of the type we seek could be constructed with this gauge group.  However, to be conservative, we focus on a model where a new SU(2)$_D$ is spontaneously broken.  For an SU(N) symmetry with a fundamental representation of dimension $N$, the $\mathcal{Z}_N$ center is preserved when the group is broken with an adjoint scalar.  To see this, note any general representation with $n_1$ upper indices and $n_2$ lower indices transforms non-trivially under the $\mathcal{Z}_N$ if $n_1 - n_2 \neq 0$.  Hence, for the SU(2)$_D$ model, the resulting $\mathcal{Z}_2$ symmetry stabilizes matter in the fundamental representation.  
\newline
\newline
\underline{A Minimal Model}:
\newline
\newline
\indent
We want to spontaneously break a gauged SU(2)$_D$ that commutes with the SM.  To do this we employ two SU(2) triplets that a singlets under SM
\begin{align}
\phi = \begin{pmatrix} \phi_2 \\ \phi_0 \\ \phi_1 \end{pmatrix} && \eta = \begin{pmatrix} \eta_0 \\ \eta_1 \\ \eta_2 \end{pmatrix}.
\end{align}  
and write down all of the relevant terms in the scalar potential that are consistent with the symmetries
\begin{eqnarray}
V &=& -{1 \over 2} m_1^2\, \phi^2 + \kappa_1 \,\phi^4 - {1 \over 2} \,m_2^2\, \eta^2 + \kappa_2 \,\eta^4  \label{eq:origpotential}  \\
&-& m_h^2\, h^\dagger h + \kappa_h \,(h^\dagger h)^2  + \kappa_3\, \phi^2 \,\eta^2 + \kappa_4\,\phi^2 \,h^\dagger h  \nonumber \\
&+& \phi^3 + \eta^3 + \kappa_5\,\eta^2 \,h^\dagger h + \kappa_6\, \eta^2\,(\phi \cdot \eta) +  \kappa_7\, \phi^2\,(\phi \cdot \eta) \nonumber \\
&+& \kappa_8\,(\phi \cdot \eta)\,h^\dagger h +  \kappa_9\, (\phi \cdot \eta)^2 + m_3^2 \,(\phi \cdot \eta) + m^4. \nonumber 
\end{eqnarray}
Here $h$ is the SM higgs which transforms under SU(2)$_L$ and hypercharge.  We have also added explicit symmetry breaking terms proportional to $\phi \cdot \eta$ to break the enhanced global SU(2)$_1$ $\times$ SU(2)$_2$ to the diagonal.  We gauge diagonal to be SU(2)$_D$.  In order to break the gauge symmetry to $\mathcal{Z}_2$ center, we want each triplet to separately break the SU(2)$_D$ to two different U(1) subgroups.  The triplet vacuum expectation values (vevs) can be
\begin{align}
\phi = \begin{pmatrix}  0 \\ v_1 \\ 0 \end{pmatrix} && \eta = \begin{pmatrix} v_2  \\ 0 \\ 0 \end{pmatrix}.
\label{eq:vevs}
\end{align} 
For the vacuum to have the proper alignment, we redefine the couplings, $\kappa_i$,  and masses, $m_i$, to generate the scalar potential
\begin{eqnarray}
V &=& \lambda_1 \biggl( \phi^2 + \lambda_6\,\phi \cdot \eta + \lambda_7\,h^\dagger h  - v_1^2 - \lambda_7\,v_h^2 \biggr)^2 \label{eq:potential}  \\           
&+& \lambda_2 \biggl(\eta^2 +  \lambda_8\,\phi \cdot \eta + \lambda_9\,h^\dagger h - v_2^2 - \lambda_9\,v_h^2\biggr)^2 \nonumber \\
&+& \lambda_3 \biggl(h^\dagger h + \lambda_{10}\,\phi \cdot \eta -v_h^2 \biggr)^2 + \lambda_4  (\phi \cdot \eta)^2 \nonumber \\
&+& \lambda_5 \biggl( \phi^2 + \eta^2 - v_1^2 - v_2^2 \biggr)^2. \nonumber
\end{eqnarray}
It is clear the vacuum prefers $\phi \cdot \eta = 0$ which is required by equation~\ref{eq:vevs}.  Of the six new degrees of freedom introduced by the triplets, three must be ``eaten" so the new SU(2)$_D$ gauge bosons, $A_i$, will get mass.  The rest ($\rho_{1,2,3}$) will get masses from spontaneous symmetry breaking.  Unitary gauge generates 
\begin{align}
 \phi = \begin{pmatrix}  {v_2 \over \sqrt{v_1^2 + v_2^2}} \,\rho_3 \\ v_1 + \rho_1 \\   \hspace{0.1cm}  \end{pmatrix} && \eta =  \begin{pmatrix} v_2 + \rho_2  \\ {v_1 \over \sqrt{v_1^2 + v_2^2}} \,\rho_3 \\ \hspace{0.1cm}  \end{pmatrix}.
\end{align}
Note there are no $v_{1,2}\, \partial \rho_3 A_i $ terms in unitary gauge.  For simplicity, we take the limit of $\lambda_i = \lambda$ and $v_{1,2} = v$ in order to outline some properties of the model.

The three new higgses mix with the SM higgs, $h_0$.  To $\mathcal{O}(\lambda)$, the charge eigenstates in terms of the mass eigenstates higgses ($h_{1,2,3,4}$) are
\begin{eqnarray}
\rho_1 &=& {1 \over \sqrt{2}}\biggl(-h_4 +  h_3 \biggr) - \lambda \biggl({\sqrt{2} \over 5} h_2 - \theta_1 h_1 \biggr)  \label{eq:rho1} \\
\rho_2 &=& {1 \over \sqrt{2}}\biggl(h_4 +  h_3 \biggr) - \lambda \biggl({\sqrt{2} \over 5} h_2 -  \theta_1 h_1 \biggr)  \label{eq:rho2}\\
\rho_3 &=& h_2 + \lambda \biggl({2 \over 5} h_3 + {\theta_2 \over \sqrt{2}}h_1 \biggr) \label{eq:rho3} \\
h_0 &=& h_1 + \lambda \biggl(\sqrt{2}\, \theta_1 h_3 + \theta_3 h_2 \biggr).
\label{eq:h0}
\end{eqnarray}
where $\theta_1 = v v_h/(3 v^2 - v_h^2)$ and $\theta_2 = v v_h/(v^2/2 - v_h^2)$.  It is clear the new higgses have parametrically suppressed couplings to the SM via the SM higgs.  After diagonalizing the mass matrix, to $\mathcal{O}(\lambda)$ the masses are $m_1 = 8 \lambda v_h^2$, $m_2 = 4 \lambda v^2$, $m_3 = 24 \lambda v^2$ and $m_4 = 8 \lambda v^2$.  We take $v > v_h$ to ensure the SM higgs, $h_1$, is the lightest higgs boson.  We can now easily show the natural correlation between the higgs and triplet vevs by simplifying the redefined parameters in equation~\ref{eq:potential}. 
\begin{eqnarray}
v^2 &=& v_{1,2}^2 = {1 \over 6 \lambda + 8 \lambda^3}\biggl(m^2 (1 + 2 \lambda^2) - 2 \lambda \,m_h^2 \biggr) \label{eq:v} \\
v_h^2 &=& {1 \over 3 \lambda + 4 \lambda^3}\biggl(3 m_h^2 - \lambda\,m^2\biggr) \label{eq:vh}
\end{eqnarray}
where $m_1 = m_2 = m$ and $m_h$ are the masses of the triplet and SM higgses, respectively, from equation~\ref{eq:origpotential}.  

The new gauge bosons become massive due to the spontaneous breaking.  The masses are 
\begin{eqnarray}
m_{A_1} &=&  g v_1  \to  m_{A_l} \equiv g v, \\
m_{A_2} &=&  g v_2  \to   m_{A_l} \equiv g v , \\
m_{A_3} &=&  g \sqrt{v_1^2 + v_2^2} \to m_{A_h} \equiv \sqrt{2} g v.
\label{eq:Amass}
\end{eqnarray}
Of interest is the $A_3$ gauge boson coupling
\begin{eqnarray}
\mathcal{L} &\supset&  {g \,v_2 A_3 \over \sqrt{v_1^2 + v_2^2}} \biggl(\rho_3 \,\partial \rho_1 - \rho_1\, \partial \rho_3 \biggr) \\
&+& {g \,v_1 A_3 \over  \sqrt{v_1^2 + v_2^2}} \biggl(\rho_2 \,\partial \rho_3 - \rho_3\, \partial \rho_2 \biggr)\nonumber \\
&\to& {g \,A_3 \over \sqrt{2}} \biggl(\rho_3 \,\partial (\rho_1 - \rho_2) - (\rho_1- \rho_2)\, \partial \rho_3 \biggr) 
\label{eq:scalar-gauge}
\end{eqnarray}
Only the $A_3$ has a linear coupling with the higgses.  All of the bosons couple quadratically to the higgses;  the $A_{1,2}$ bosons decay at the two loop level to the SM gauge bosons and higgses. 

We now add fermions doublets which transform non-trivially under the $\mathcal{Z}_2$ center.  In a companion paper~\cite{Walker:2009ph}, we show how long-lived particles can be used at the LHC to experimentally differentiate different dark matter stabilization symmetries.  To provide support, we add a minimal, anomaly free content of fermions transforming as
\begin{eqnarray}
\psi  &=& (3,1,2)_{-2/3} \oplus (1,2,2)_{1} \\
 &\oplus& (\overline{3},1,2)_{2/3} \oplus (1,2,2)_{-1}. \nonumber
\label{eq:fermioncontent}
\end{eqnarray}
Here the entries in parenthesis are color, SU(2)$_L$ and SU(2)$_D$ representations.  The subscripts are the hypercharge assignments.  We choose fermions that are vector-like to reduce the sensitivity to precision electroweak measurements.  A minimal choice for the dark matter candidates is
\begin{equation}
\chi = (1,2) \oplus (1,\overline{2})
\label{eq:darkmatter}
\end{equation}
The yukawa couplings are
\begin{eqnarray}
\mathcal{L}_\mathrm{yukawa} &=& \lambda_6 \,\psi^* \phi\, \psi + \lambda_7 \,\psi^* \eta\, \psi \\
&+& \lambda_8 \,\chi^* \phi\, \chi + \lambda_9 \,\chi^* \eta\, \chi \nonumber
\end{eqnarray}
where $\eta$ and $\phi$ have the SU(2)$_D$ fundamental indicies.  The mass terms for both particles are
\begin{eqnarray}
m_\psi = \sqrt{\lambda_6^2 v_1^2 + \lambda_7^2 v_2^2} \to \sqrt{2} \,\lambda_\psi v\\
m_\chi = \sqrt{\lambda_8^2 v_1^2 + \lambda_9^2 v_2^2} \to \sqrt{2} \,\lambda_\chi v.
\end{eqnarray}
We assume the fermions get their mass solely from SU(2)$_D$ spontaneous symmetry breaking.  We took the simplifying limit of $\lambda_6 = \lambda_7 = \lambda_\psi$ and $\lambda_8 = \lambda_9 = \lambda_\chi$.  We also assume $\lambda_\chi < \lambda_\phi$ so the dark matter candidates are $\chi$ fermions.  After spontaneous symmetry breaking $\psi$ and $\chi$ are no longer SU(2)$_D$ doublets; thus, quantum correction can potentially lift the degeneracy between the fermions in the multiplet.

We have not provided a mechanism for the long-lived heavy quarks and leptons in equation~\ref{eq:fermioncontent} to decay despite strong constraints~\cite{Perl:2001xi}.  To be consistent with these bounds, we can posit a scenario where SU(2)$_D$ is unified with SU(2)$_L$ into a simple gauge group, $\mathcal{G}$~\cite{dawsongeorgi}.  Breaking $\mathcal{G}$ generates new gauge bosons of mass, $\Lambda$, which generate the effective operators 
\begin{equation}
\mathcal{O}_2 = {1 \over \Lambda^2} \,q \,q \,\psi \chi.
\end{equation}
The heavier $\Psi$ fermion would then decay to the SM and $\chi$ with a lifetime, $\tau \sim \Lambda^4/m_\psi^5$.
\newline
\newline
\underline{Precision Electroweak}:
\newline
\newline
All the new fermions, gauge and higgs bosons introduced in this model interact with the SM by mixing with with the SM higgs (see equation~\ref{eq:h0}).  Because of the connection through the SM higgs, the most important precision electroweak corrections come from large custodial SU(2) violations.  The corrections to the low energy effective theory are quantified by the effective operator~\cite{precisionew} 
\begin{equation}
\mathcal{O}_3 = {c \over M^2}  h^\dagger D_\mu h\,\,h^\dagger D_\mu h
\label{custSU2}
\end{equation}
where M is the mass of the new physics generating the effective operator and $D_\mu$ is the electroweak covariant derivative.  From presentation above, it is clear there are no custodial SU(2) violating couplings which would generate tree level corrections to equation~\ref{custSU2}.  The other precision electroweak measurements are only mildly sensitive to deviations from the SM higgs properties.
\newline
\newline
\underline{Relic Density}:
\newline
\newline
\indent
With minimal constraints from precision electroweak measurements on the model parameter space, we calculate the thermal relic abundance needed to recover the measured relic density of  $h^2\, \Omega_{\mathrm{DM}} = 0.1131 \pm 0.0034$ \cite{Hinshaw:2008kr}.  The contribution to the density from a non-relativistic species is 
\begin{equation}
\Omega_\chi \approx {1.04 \times 10^9 \over M_\mathrm{pl}} {x_F \over \sqrt{g_*}} {1 \over (a + 3 b/x_F)}
\end{equation}
where $x_F = m_\chi/T_F$, $g_*$ counts the number of relativistic degrees of freedom and $T_F$ is the freeze out temperature.  The coefficients $a$ and $b$ are computed from the annihilation cross section in the low velocity limit.
\begin{equation}
\langle\sigma v\rangle = a + b\,  \langle v^2\rangle + \mathcal{O}(\langle v^4\rangle) 
\end{equation}
Details for calculating the abundances can be found in \cite{relic}.  The $\chi_i$ dark matter, dominantly annihilates via $\overline{\chi}_i \,\chi_i \to A_3$; $A_3$ subsequently annihilates into all the kinematically allowed higgses in equation~\ref{eq:scalar-gauge}.  
Note, the triplet vev effectively sets the mass scale for all the particles involved in the dark matter annihilation process.  In figure 1, we perform a parameter scan over the dark matter mass and the triplet vev $v$ for all the values which satisfy the measured relic abundance.  The mass of the new heaviest gauge boson, $A_h$, is given in equation~\ref{eq:Amass}.  We set the SM higgs mass and vev, $v_h$, to 120 and 246 GeV, respectively.  This in turn specifies $\lambda$ and the higgs mass hierarchy.
\begin{figure}
\label{fig:relic1}
{\includegraphics[width=7.0truecm,height=5truecm,clip=true]{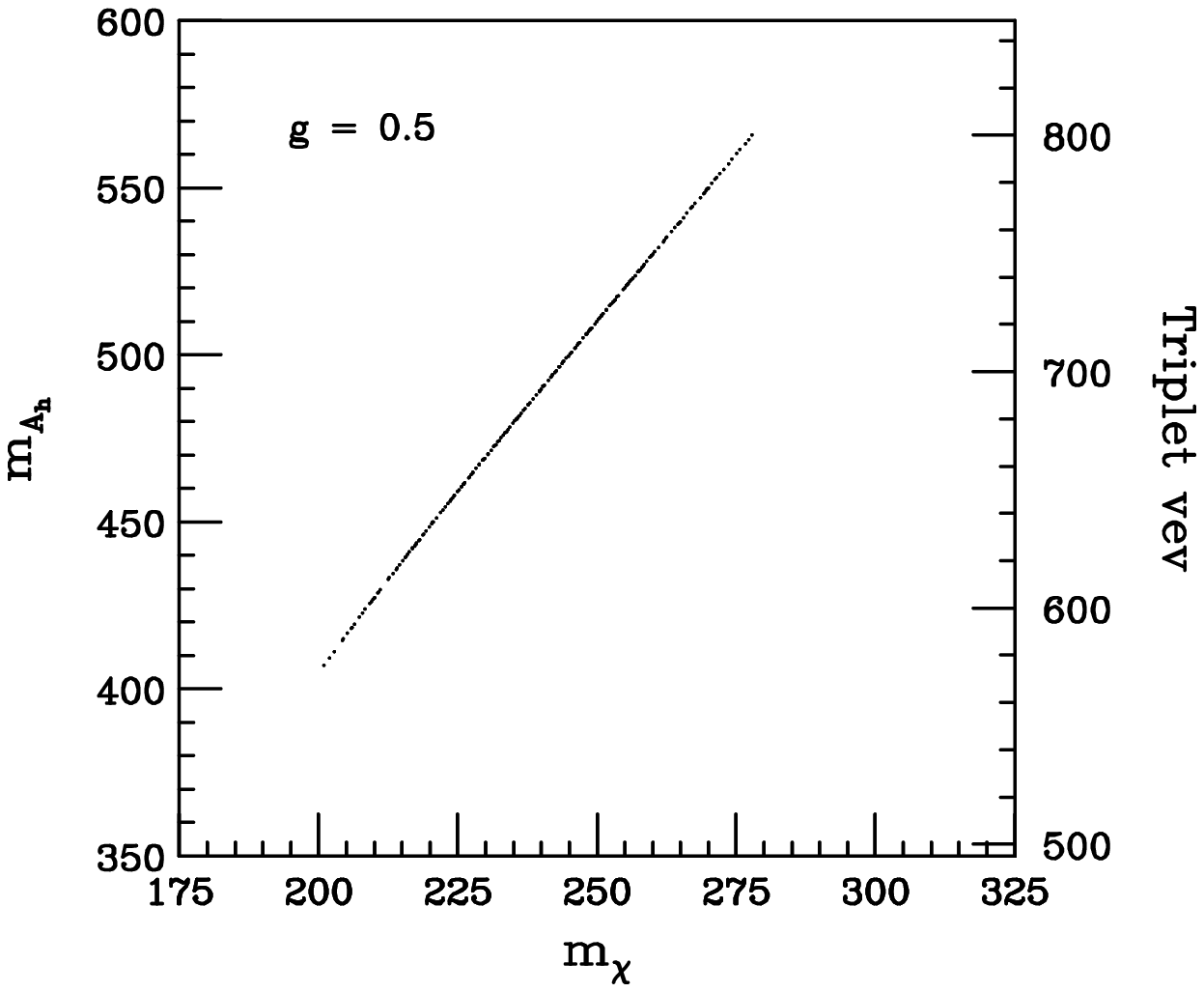} \newline
\includegraphics[width=7.0truecm,height=5truecm,clip=true]{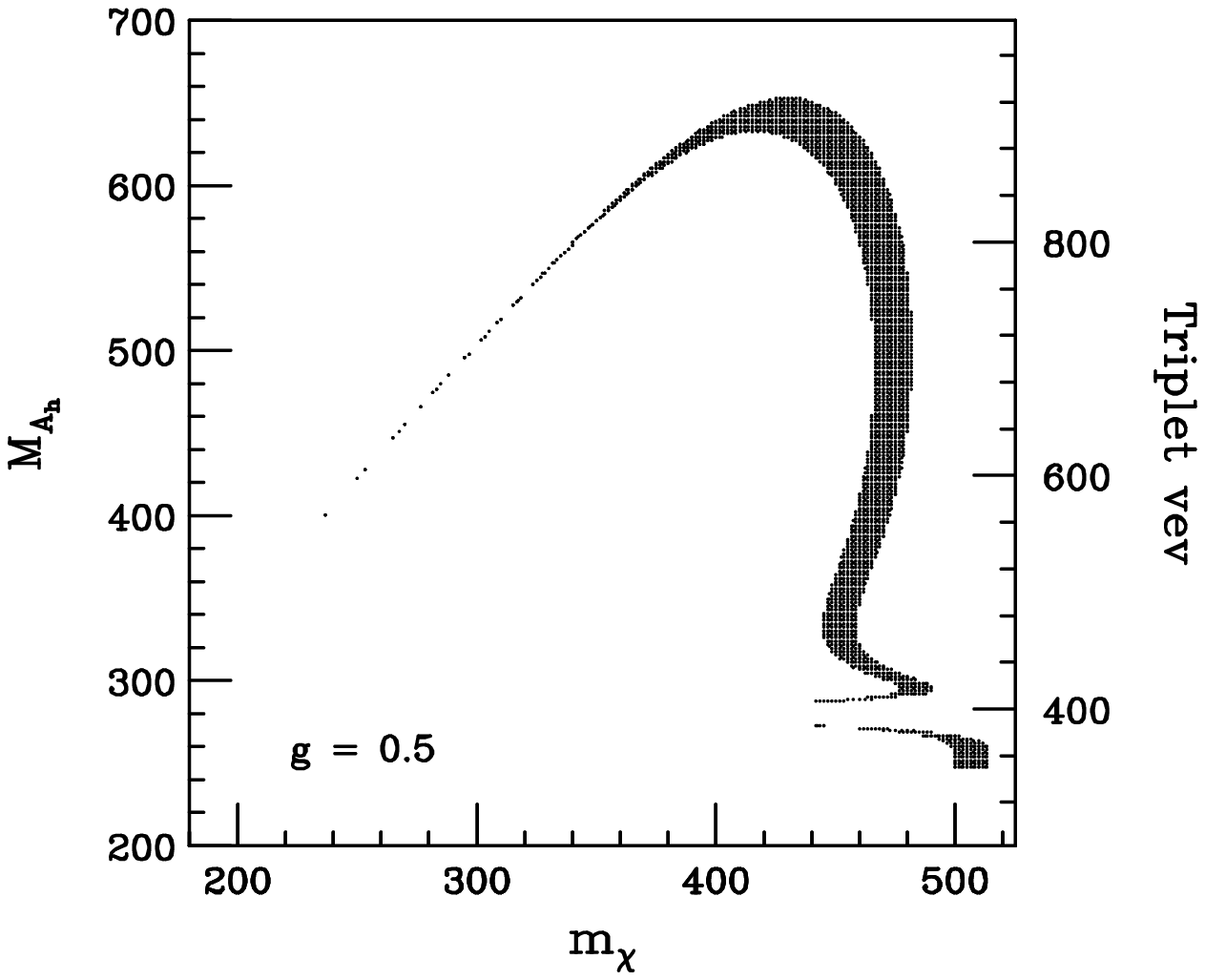} \newline}
\caption{Parameter scan of the dark matter mass versus the mediating particle gauge boson mass for a new gauge coupling of $g = 0.5$.  The higgs mass hierarchy is set by requiring the lightest/SM higgs  mass, $m_h = 120$ GeV.  In the top panel, the parameter space for the $\chi$ dark matter to annihilate into the lightest two higgs bosons is shown.  The bottom panel shows the parameter space for the dark matter to annihilate to the lightest two sets of higgses.}
\end{figure}
\newline
\newline
\underline{LHC Signatures}:
\newline
\newline
\indent
Strong evidence for this class of model comes from the interactions of the new gauge bosons.  As shown above, the $A_3$ boson can decay directly to kinematically favorable higgses; but, the $A_{1,2}$ bosons decay directly into the dark fermions when $m_\chi < m_{A_i}$.  A companion paper~\cite{Walker:2009ph} demonstrates how signatures with long-lived particles and a large amount of missing transverse energy ($\etmiss$) can distinguish models with dark matter stabilized with a $\mathcal{Z}_2$ versus other symmetries at the LHC.  The current model generates this missing energy by radiating off the $A_{1,2}$ gauge bosons from the long-lived $\psi$ fermions.  These bosons dominantly decay into the $\chi$ dark matter with almost a 100\% branching fraction.  The $A_3$ boson, additionally, can decay into higgses with a significant branching fraction.  If the higgses are lighter than twice the Z mass, they can subsequently decay into b quarks.  The final signature would be $pp \to \overline{\psi}\, \psi \,+$ 4 $b$ quarks.  Tagging at least one long-lived particle and at least one b quark is required.  Reconstructing nearest neighbor jets closest to the tagged b quark must reconstruct the higgs invariant mass in order to reduce the QCD background.  If the higgses are heavier than twice the Z mass, a decay into $pp \to \overline{\psi}\, \psi \,+$ 4 $Z$ bosons is possible and very clean when the $Z$ decays to leptons.  As reviewed in~\cite{Kraan:2005ji}, long-lived quarks can charge flip as they interact with valence quarks in the detector.  An additional dramatic signal is possible if one the of long-lived quarks charge flips to neutral.  A clean final state would feature events with only one long-lived particle, large $\etmiss$ and four $b$ quarks.~(or $Z$ bosons)

In all, we demonstrated a model where the dark matter stabilization symmetry is generated by spontaneous symmetry breaking.  This model naturally ties the weak scale with the scale associated with the dark matter annihilation cross section.  We have also shown the thermal relic abundance for the simplest models as well as sketched distinct LHC signatures.

{\it Acknowledgments:}
We thank A. E.~Nelson, M.~Schmaltz and T.~Tait for useful conversations.  This work was supported by a University of California Presidential fellowship.


\begin{thebibliography}{000}

\bibitem{Hinshaw:2008kr}
  G.~Hinshaw {\it et al.}  [WMAP Collaboration],
  arXiv:0803.0732 [astro-ph].

\bibitem{Kolb:1990vq}
  E.~W.~Kolb and M.~S.~Turner,
  Front.\ Phys.\  {\bf 69}, 1 (1990).
  
\bibitem{Feng:2008ya}
  J.~L.~Feng and J.~Kumar,
  Phys.\ Rev.\ Lett.\  {\bf 101}, 231301 (2008).

\bibitem{Walker:2009ph}
  D. G. E.~Walker,
  arXiv:0907.3142 [hep-ph]. 

\bibitem{agashe:2009ph}
K.~Agashe, D.~Kim, N.~Moore, M. Toharia and D. G. E.~Walker,
in progress.
    
\bibitem{Bertone:2004pz}
Please see the following review for a survey of popular models.
  G.~Bertone, D.~Hooper and J.~Silk,
  Phys.\ Rept.\  {\bf 405}, 279 (2005)
  [arXiv:hep-ph/0404175].
     
\bibitem{abelian}
 See, for example, E.~G.~Adelberger, B.~R.~Heckel and A.~E.~Nelson,
  Ann.\ Rev.\ Nucl.\ Part.\ Sci.\  {\bf 53}, 77 (2003)
  [arXiv:hep-ph/0307284];
  C.~Amsler {\it et al.}  [Particle Data Group],
  Phys.\ Lett.\  B {\bf 667}, 1 (2008);
  S.~Davidson, S.~Hannestad and G.~Raffelt,
  JHEP {\bf 0005}, 003 (2000);
  B.~Holdom,
  Phys.\ Lett.\  B {\bf 166}, 196 (1986).

\bibitem{georgi:2009ph2}
 H.~Georgi, T.~Tait,  D. G. E.~Walker, and C.-P.~Yuan, in progress.
 
\bibitem{Perl:2001xi}
  M.~L.~Perl, P.~C.~Kim, V.~Halyo, E.~R.~Lee, I.~T.~Lee, D.~Loomba and K.~S.~Lackner,
  Int.\ J.\ Mod.\ Phys.\  A {\bf 16}, 2137 (2001)
  [arXiv:hep-ex/0102033].

\bibitem{dawsongeorgi}
  S.~Dawson and H.~Georgi,
  Phys.\ Rev.\ Lett.\  {\bf 43}, 821 (1979).
  
\bibitem{precisionew}
  M.~E.~Peskin and T.~Takeuchi,
  Phys.\ Rev.\  D {\bf 46}, 381 (1992);
  B.~Grinstein and M.~B.~Wise,
  Phys.\ Lett.\  B {\bf 265}, 326 (1991).

  
\bibitem{relic}
  G.~Jungman, M.~Kamionkowski and K.~Griest,
  Phys.\ Rept.\  {\bf 267}, 195 (1996);
  G.~Servant and T.~M.~P.~Tait,
  Nucl.\ Phys.\  B {\bf 650}, 391 (2003).

\bibitem{Kraan:2005ji} 
  A.~C.~Kraan, J.~B.~Hansen and P.~Nevski,
  Eur.\ Phys.\ J.\  C {\bf 49}, 623 (2007).
 \end{thebibliography}
\end{document}